# Magnetic anisotropy and magnetoresistance of sputtered [(FeTaN)/(TaN)]$_n$ multilayers

H. B. Nie[1], S. Y. Xu [a),1], J. Li[1], C. K. Ong[1] and J. P. Wang[2]

[1] *Centre for Superconducting and Magnetic Materials, Institute of Engineering Science and Department of Physics, National University of Singapore, Science Drive 3, Singapore 117542;*

[2] *Data Storage Institute, DSI Building, 5 Engineering Drive 1, Singapore 117608*

We studied the in-plane magnetic anisotropy of RF (radio frequency) sputtered [(FeTaN)/(TaN)]$_n$ multilayers synthesized on Si substrates. In the multilayers where n = 5, the FeTaN thickness is fixed at 30 nm and the thickness of TaN, $t_{TaN}$, is varied from 0 to 6.0 nm, we observed a clear trend that, with increasing $t_{TaN}$, the values of coercivity, grain size and amplitude of maximum magnetoresistance (MR) of the samples all decrease first and then increase after reaching a minimum when $t_{TaN}$ is around 2.0-4.0 nm. This trend is also associated with an evolution of in-plane magnetic anisotropy, where the multilayers change from uniaxial anisotropy to biaxial at $t_{TaN}$ around 4.0 nm and above. We attribute the phenomena to the interlayer coupling effect of FeTaN films as a function of the coupling layer (TaN) thickness, rather than to the thickness dependence observed in single-layered FeTaN films, where the direction of easy axis switches 90º when the film is thicker than 300 nm. The in-plane anisotropy of the [(FeTaN)/(TaN)]$_n$ multilayers also shows signs of oscillation when the number of coupling layers varies. The MR effects observed are mainly due to anisotropy MR (AMR), while the grain size and exchange coupling may also contribute to the change of maximum MR ratios in the multilayers with changing $t_{TaN}$.

## I. INTRODUCTION

As one of the potential candidates for the novel write-head application in data storage and processing systems, iron-based nitride multilayers have attracted much attention in recent years.[1-3] Compared to single-layered iron-based films, iron-based multilayers have the advantages of low thermal noise, low loss, high power-handling capacity and thermal stability for high frequency performance up to the order of 100 MHz that fulfill the strict requirements for ultrahigh density recording. We synthesized and systematically investigated one of the iron-based multilayers, [(FeTaN)/(TaN)]$_n$, a system that has been intensively studied.[4-6] In this report, we mainly present our results on properties of the in-plane anisotropy related to interlayer coupling, and the in-plane magnetoresistance (MR) of the multilayers. The studies may have both scientific and technical importance.[7-8]

## II. EXPERIMENT

The films were synthesized on Si(100) substrates by using reactive RF magnetron sputtering on a Denton Vacuum Discovery-18 Deposition System. A 3-inch Fe target with Ta chips that cover about 2.0-3.0 % of the efficient Fe target surface area was used for FeTaN films. An aligning magnetic field $H_{al}$ of 150 Oe was applied to the substrate, parallel to the substrate surface. The parameters used for film deposition were: the base vacuum 8.0 x $10^{-7}$ torr, total ambient N2/Ar pressure 2.0 mtorr, bias voltage –100 V. The percentage of $N_2$ partial pressure $P_N$ in the N2/Ar ambient gasses was varied for different films. For FeTaN films, $P_N$ was set at 3.0% and the RF power was kept at 100 W. For TaN films, $P_N$ was set at 20.0% and the RF power was kept at 70 W, while the resistivity of the TaN films was around 1.13 mΩ·cm. In the series of $[(FeTaN)/(TaN)]_5$ multilayers, the thickness of FeTaN layers was kept at 30±3 nm, while the thickness of TaN interlayer, $t_{TaN}$, was varied from 0 nm to 6.0 nm, by a step of 1.0 nm. In addition, $[(FeTaN)/(TaN)]_n$ multilayers with varied period number n=1-8 were also made for comparison. We characterized the as-deposited samples using X-ray diffraction (XRD), atomic and magnetic force microscopy (AFM and MFM), angle-resolved M-H loop tracer, vibration sample spectrometer (VSM) and four-probe measurement method to study their structural and magnetic properties as well as MR ratios.

## III. RESULTS AND DISCUSSION

Previously we had studied FeN thin films synthesized with the same sputtering system described above [9]. In 30-nm-thick FeN thin films, their easy axes are all lying along the direction of aligning magnetic field $H_{al}$. With increasing of $P_N$, the coercivity $H_c$ first decreased, then increased, and had minimum values of 2-3 Oe when $P_N$ ranges between 3.0% and 7.0%. In FeN thin films, with the increase of $P_N$, from 0.0% to 12.0%, the easy-axis squareness $S_e$ ($S = M_r/M_s$) of the in-plane M-H hysteresis loops is almost a constant but the hard-axis squareness $S_d$ shows a trend similar to $H_c$ and minimizes at 0.18 when $P_N$ = 7.0%. While in FeTaN thin films, both $S_e$ and $S_d$ do not change much with $P_N$ varying in the same range. Here Ta atoms in FeTaN seem to make the lattice structure more stable. However, in FeTaN films, we have noted the dependence of in-plane $H_c$ with film thickness. In thin films, the film is anisotropic with its easy axis lying along $H_{al}$. With increase of film thickness, the film becomes isotropic and then its easy axis appears along 90º to $H_{al}$, as shown in Fig. 1. The underlying mechanism, which could relate to structure, stress distribution and grain size, is not clear.

The coercivity of $[(FeTaN)/(TaN)]_5$ multilayers shows a thickness dependence on $t_{TaN}$ (Fig. 2(a)). Calculated from XRD step-scan patterns and AFM micrographs, the average grain size of the samples, ranging between 20 and 30 nm (Fig. 2(b)), and the root-mean square roughness Rq, ranging between 0.3 and 0.6 nm, both have the same trends as that of $H_c$, i.e., with increasing $t_{TaN}$, they decrease first, reach a minimum at $t_{TaN}$ = 2.0-3.0 nm and then increase. As revealed by XRD, the dominating crystalline component of the multilayers is nano-sized crystalline grains of *bcc* α-Fe, which show only (110) peak in the XRD patterns. With increasing $t_{TaN}$, the corresponding lattice spacing $d_{(110)}$ first slightly increases, then decreases (Fig. 2 (d)). One can expect that, the change of $d_{(110)}$ could induce a change in the stress distribution in the multilayer thus changes in the properties of the sample such as $H_c$, magnetic anisotropy and MR. The trends of $H_c$, grain size, $R_q$ and $d_{(110)}$ changing with $t_{TaN}$ are well consistent, therefore a common mechanism may exist accounting for the similar trends. The grain size could play a key role. In sputtered films, grain size increases with film thickness. In the periodical $[(FeTaN)/(TaN)]_n$ multilayers, reduces the thickness of interlayer TaN causes an interruption of the growth of FeTaN layer thus decreases its grain size. The effect works until the thickness dependence of grain size dominates the film growth again. However, it is unexpected that the grain size of [(FeTaN)/(TaN)]$_5$ is so sensitive to $t_{TaN}$. In our samples, the minimum of $H_c$ at $t_{TaN}$ = 2.0-3.0 nm is a natural result from the enhanced exchange coupling between grains with minimum grain size, [10-11] which is around 20 nm at the same $t_{TaN}$, as well as from the reduction of domain-wall energy due to magnetostatic coupling of FeTaN layers across TaN interlayer. [10, 12-13]

We did not observe perpendicular magnetic anisotropy in the [(FeTaN)/(TaN)]$_5$ multilayers. However, the angle dependence of in-plane coercivity of the samples presents an evolution of anisotropy from isotropy when $t_{TaN}$ = 0, i.e. a 150-nm-thick FeTaN layer, to uniaxial anisotropy along the direction of $H_{al}$ at $t_{TaN}$ = 1.0 nm, and to biaxial anisotropy when $t_{TaN}$ is 4.0 nm or above (Fig. 3). The biaxial anisotropy could not be explained as a combination of two uniaxial anisotropies that have 90º to each other as observed in FeTaN films (Fig. 1). In the [(FeTaN)/(TaN)]$_5$ multilayers, the sum thickness of FeTaN layers adds up to only 150 nm, at which, the single-layered FeTaN film still shows uniaxial anisotropy along $H_{al}$. It means that the appearance of biaxial anisotropy should be induced by interlayer magnetic coupling rather than by the thickness effect. It is further verified by the results taken from the [(FeTaN, 30nm)/(TaN, 4.0nm)]$_n$ samples, where the period number n was varied from 1 to 8. In these samples, the biaxial anisotropy is obviously observed when n = 3, 5 and 8, but not obvious in the rest, showing the existence of an oscillation on period-number that needs further studies. [14]

The [(FeTaN)/(TaN)]$_5$ multilayers have typical metallic behavior with temperature. The values of their in-plane resistivity increase with increasing $t_{TaN}$ as TaN has much higher resistivity than that of $\alpha$-Fe, [15] and more grain boundary scattering and impurity scattering is induced with increasing $t_{TaN}$. Fig. 4 shows a typical in-plane MR pattern of the sample of [(FeTaN, 30 nm)/(TaN, 3.0 nm)]$_5$, where the measuring bias current $I$ and external magnetic field $H$ are both applied in the film plane. Here the MR ratio is defined as $\Delta\rho(H) = [\rho(H)-\rho_s]/\rho_s$, where $\rho(H)$ is the sample resistivity at field $H$ and $\rho_s$ is the resistivity at $H_s$ = 28.5 Oe. When $H$ is perpendicular to the bias current $I$, the films show positive MR; when $H$ is parallel to $I$, the films show negative MR. At $H = \pm H_c$, the values of MR have maximum amplitudes. Obviously it is a typical anisotropic magnetoresistance (AMR) effect that can be well described with the Voigt-Thomson formula $\rho = \rho_0 + \Delta\rho \cos^2\theta$. The maximum MR ratios (taken as an average of the two values measured at $H = \pm H_c$) of the samples also change with $t_{TaN}$ (Fig. 2(c)), having a similar trend to that of $H_c$ and grain size, either at room temperature (RT) or at liquid nitrogen temperature (LT). It is natural that films with smaller grains, magnetic exchange coupling of the grains is enhanced and it results in a more uniform distribution of magnetization, thus a smaller MR. The second factor is that, with the decrease of grain size, the grain boundary and interfacial scatterings are remarkably enhanced, which result in a higher resistivity $\rho(0)$ as well as $\rho_s$ of the film. Suppose that $\rho(H)-\rho_s$ keeps the same value, then MR is reduced. In addition, we did not observe stripe domains in the [(FeTaN)/(TaN)]$_5$ samples, therefore the contribution of giant magnetoresistance (GMR) effect due to strip domains [16] could be neglected.

## IV. CONCLUSION

We have shown that [(FeTaN)/(TaN)]$_5$ multilayers have consistent dependence of coercivity, grain size and MR ratio with the interlayer thickness $t_{TaN}$, which seems dominated by grain size effect and interlayer coupling. The appearance of biaxial in-plane anisotropy of the multilayers at $t_{TaN}$ around 4.0 nm is attributed to interlayer coupling effect, rather than the thickness effect observed in FeTaN films where the easy axis switches 90º in thick films. The in-plane MR behavior observed in the multilayers is mainly resulted from AMR effect, while grain-size-related exchange coupling, grain boundary and interfacial scattering effects may also contribute to the MR dependence on $t_{TaN}$. Signs of oscillation of in-plane anisotropy with the periodical number of TaN were observed and it needs further study to reveal the mechanism.

**References**

[1] Nobuyuki Ishiwata, Chizuko Wakabayashi and Haruo Urai, J. Appl. Phys. **69**, 5616 (1991).
[2] G. Qiu, E. Haftek, and J. A. Barnard, J. Appl. Phys. **73**, 6573 (1993).
[3] P. Zheng, J. A. Bain and M. H. Kryder, J. Appl. Phys. **81**, 4495 (1997).
[4] M. Naoe and S. Nakagawa, J. Appl. Phys. **79**, 5015 (1996).
[5] S. X. Li, P. P. Freitas, M. S. Rogalski, M. Azevedo, J. B. Sousa, Z. N. Dai, J. C. Soares, N. Matsakawa and H. Sakakima, J. Appl. Phys. **81**, 4501 (1997); S. X. Li, P. P. Freitas, S. Cardoso, J. C. Soares, B. Almeida, J. B. Sousa, J. Magn. Magn. Mater. **165**, 363 (1997).
[6] M. S. Rogalski, M. M. Amado, J. B. Sousa, P. P. Freitas, J. Magn. Magn. Mater. **197,** 75 (1999).
[7] P. Grünberg, Acta Mater. **48**, 239 (2000).
[8] J. L. Costa-Krämer, J. L. Menéndez, A. Cebollada, F. Briones, D. García and A. Hernando, J. Magn. Magn. Mater. **210,** 341 (2000).
[9] H. B. Nie, S. Y. Xu and C. K. Ong, International Conference on Materials for Advanced Technologies, Singapore, E8-13, Abstr. P. 178, 1-6 July 2001; http://www.mrs.org.sg/icmat2001/.
[10] M. H. Kryder, S. Wang and K. Pook, J. Appl. Phys. **73**, 6212 (1993).
[11] G. Herzer, IEEE Trans. Mag. **26**, 1397 (1990); J. Mat. Eng. Perform. **2**, 193 (1993).
[12] H. Clow, Nature **194**, 1035 (1962).
[13] J. C. Slonczewski and S. Middlehoek, Appl. Phys. Lett. **6**, 139 (1965).
[14] J. C. Slonczewski, J. Magn. Magn. Mater. **150**, 13 (1995).
[15] H. B. Nie, S. Y. Xu, S. J. Wang, L. P. You, Z. Yang, C. K. Ong, J. Li and T. Y. F. Liew, Appl. Phys. A **73**, 229 (2001).
[16] H. S. Cho, V. R. Inturi, J. A. Barnard and H. Fujiwara, IEEE Trans. Magn. **34**, 1150 (1998).

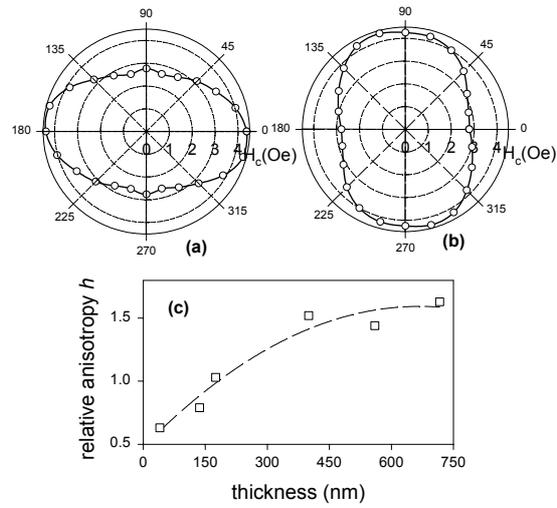

Figures 1. (a) and (b) show the typical polar plots of $H_c$ of two FeTaN films with thickness' of 40 nm and 400 nm, respectively. (c) shows the in-plane relative anisotropy $h$, which is defined as $h = [H_c(90º)+H_c(270º)]/[H_c(0º)+ H_c(180º)]$, as a function of thickness of FeTaN films deposited at $P_N = 3.0\%$.

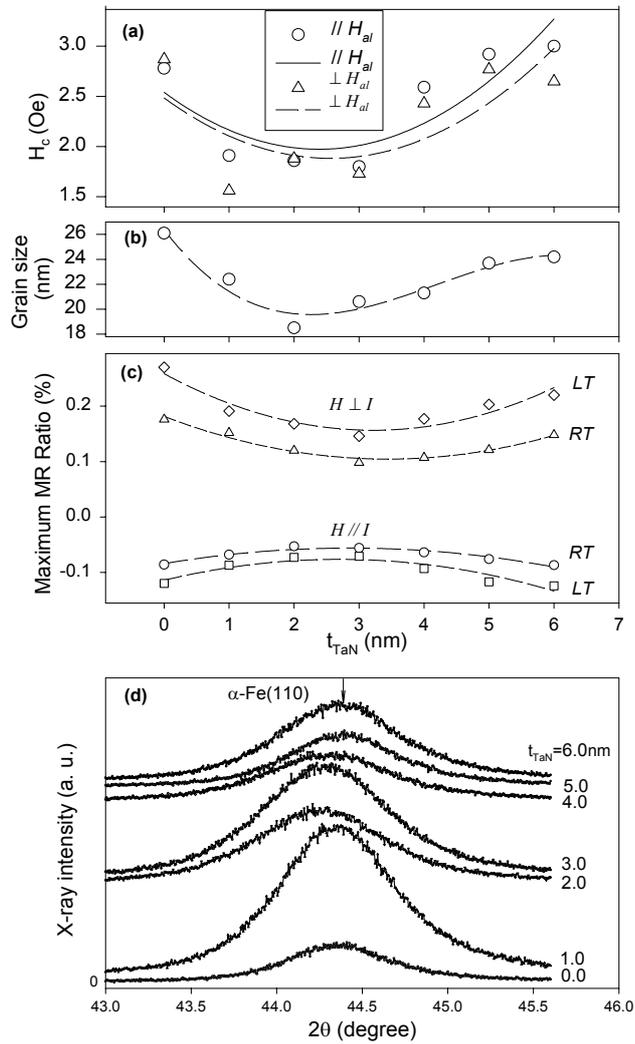

Figures 2(a), 2(b) and 2(c) show the dependences of $H_c$, grain size and maximum MR ratios of $[(FeTaN)/(TaN)]_5$ multilayers as functions of $t_{TaN}$. Figure 2(d) plots the XRD step-scan patterns of the same samples, showing a shift of (110) peak position with increase of $t_{TaN}$, corresponding to the change of $d_{(110)}$ spacing in the films. In (a), the circles are for the applied field parallel to $H_{al}$, the triangles are for the applied field perpendicular to $H_{al}$; In (b), the diamonds and the triangles are for the applied field perpendicular to current at low temperature (77.7 K) and at room temperature respectively, the circles and the squares are for the applied field parallel to current at room temperature and at low temperature (77.7 K) respectively.

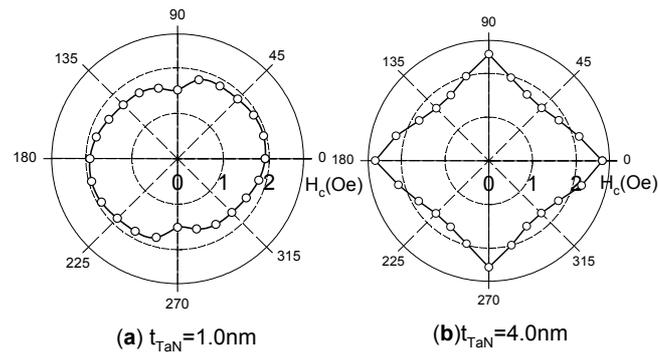

Figure 3 Plots the in-plane $H_c$ of two $[(FeTaN)/(TaN)]_5$ multilayers at the interlayer thickness $t_{TaN}$ = 1.0 nm (a) and $t_{TaN}$ = 4.0 nm (b) as a function of the angle with aligning field $H_{al}$.

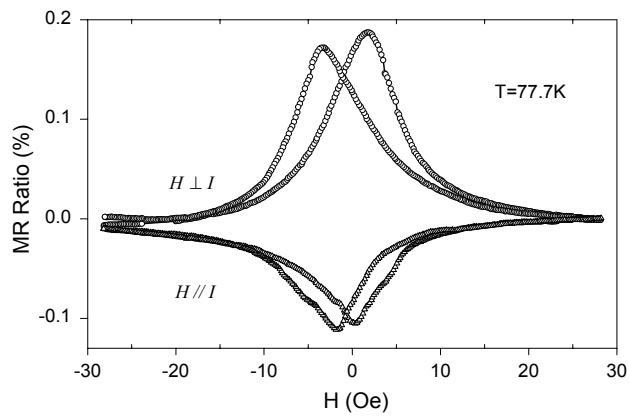

Figure 4 Typical in-plane MR hysteresis loops of a $[(FeTaN, 30\ nm)/(TaN, 3.0\ nm)]_5$ sample tested at 77.7K.